\documentclass[preprint,12pt]{elsarticle}

\usepackage{amssymb}
\usepackage{bm}
\usepackage{amsthm}

\usepackage{lineno}

\newcounter{bla}

\journal{Computer Physics Communications}

\begin{document}

\begin{frontmatter}



\title{ANEMONE: a fully three-dimensional solid-state electro-aerodynamic propulsion system simulator}


\author[a]{Hisaichi Shibata\corref{author}}
\author[a]{Soya Shimizu}
\author[b]{Takahiro Nozaki}

\cortext[author] {Corresponding author.\\\textit{E-mail address:} altair@keio.jp}
\address[a]{Graduate School of Science and Technology, Keio University, 3-14-1 Hiyoshi, Kohoku-ku, Yokohama-shi, 223-8522 Kanagawa, Japan.}
\address[b]{Department of System Design Engineering, Keio University, 3-14-1 Hiyoshi, Kohoku-ku, Yokohama-shi, 223-8522 Kanagawa, Japan.}

\begin{abstract}
Solid-state electro-aerodynamic propulsion systems are devices that utilize atmospheric pressure corona discharge and have been actively researched in recent years as a means of achieving silent drones. However, these systems contain multiple, widely disparate time and spatial scales. Therefore, the governing equations of the systems, a three-component plasma fluid model that considers the presence of electrons, positive ions, and negative ions, constitute a stiff non-linear system of partial differential equations, challenging to solve. Here, we have developed an ANEMONE simulator capable of numerically estimating the corona inception voltage and energy conversion efficiency in three-dimensional solid-state electro-aerodynamic propulsion systems. Specifically, on the basis of the governing equations, we adopted the method of characteristics and the perturbation method to obtain the sub-problems. Furthermore, we have successfully obtained the integral equations, making the sub-problems easier to solve. Finally, we validated the prediction results based on the theoretical results in a previous study. Remarkably, ANEMONE is the first simulator in the world which predicted the two representative performance of fully three-dimensional propulsion systems. 
\\


\noindent \textbf{PROGRAM SUMMARY/NEW VERSION PROGRAM SUMMARY}

\begin{small}
\noindent
{\em Program Title:} ANEMONE                                          \\
{\em CPC Library link to program files:} (to be added by Technical Editor) \\
{\em Developer's repository link:} (if available) \\
{\em Licensing provisions(please choose one):} MIT \\
{\em Programming language:} C++                                \\
{\em Supplementary material:}                                 \\
{\em Journal reference of previous version:}*                  \\
{\em Does the new version supersede the previous version?:}*   \\
{\em Reasons for the new version:*}\\
{\em Summary of revisions:}*\\
{\em Nature of problem(approx. 50-250 words):} Solid-state electro-aerodynamic propulsion system [1] ionizes the ambient air and can propel silent drones. For the rapid-prototyping of the system, it is important to utilize numerical simulators, but the spatial and temporal scales of the system are diverse; hence, the corresponding governing equations (e.g. the three-component plasma fluid model which can simultaneously consider electrons, positive and negative ions) are too stiff to solve. For example, the overall spatial scale of the system is the order of meters, while the scale of the electrodes is the order of micrometers. Moreover, the ions move between electrodes in the order of milliseconds, while the Maxwell dielectric relaxation time scale is in the order of nanoseconds.
\\
{\em Solution method(approx. 50-250 words):} For the issue of the spatial scales, we adopt a three-dimensional hierarchical Cartesian grid method together with the adaptive mesh refinement method. Moreover, this leads fully automatic mesh generation and ensures the grid convergence. For the issue of the temporal scales, we adopt the perturbation method [2] combined with the method of characteristics. This can decompose the original problem into many subproblems easy to solve.
\\
{\em Additional comments including restrictions and unusual features (approx. 50-250 words):}\\
   \\

* Items marked with an asterisk are only required for new versions
of programs previously published in the CPC Program Library.\\
\end{small}
   \end{abstract}
\end{frontmatter}
\section{Introduction}
\label{sec:intro}
Solid-state electro-aerodynamic propulsion systems are devices that utilize atmospheric pressure corona discharge and have been actively investigated in recent years as a means of achieving silent drones \cite{xu2018flight, xu2019higher, gilmore2015electrohydrodynamic, masuyama2013performance, gomez2023model, gomez2024order}. For rapid societal implementation of these devices, numerical simulators \cite{martins2011influence, martins2011modeling, martins2012simulation, martins2013modelling, picella2024numerical, shibata2016performance, shibata2022novel, xu2019higher} are essential. Such a simulator would replace high-risk, high-voltage testing which requires significant human effort and caution by enabling rapid prototyping through understanding and modeling of the underlying physical phenomena. 

However, the spatial and temporal scales of the solid-state electro-aerodynamic propulsion system are diverse. Specifically, the overall spatial scale of the devices is on the order of meters. In contrast, in the vicinity of the corona active electrode, where the radius of curvature is extremely small, it is necessary to resolve steep potential gradients and abrupt charge-density variations, generally requiring a dense grid. Given that the electrode curvature radius is 100 micrometers, a grid with at least approximately 1/10 of this resolution is considered necessary to accurately capture physical phenomena. However, setting the entire computational grid to the high density is nearly impossible due to memory limitations, even with state-of-the-art supercomputers. Therefore, a stretched computational grid is required. Automatically generating such a spatially multiscale unstructured grid for general electrode shapes and electrode arrangements with complex geometries is typically challenging.

In the aviation community, automatic grid generation and fluid computations have been performed using a hierarchical Cartesian grid method (e.g. FFVHC-ACE \cite{asada2023ffvhc}). They have achieved large-eddy simulations (LES) in practical high Reynolds number regimes by combining wall models formulated with ordinary differential equations and finite-difference (or finite-volume) methods.  
In this study, we adopt another hierarchical Cartesian grid method which enables fully automatic meshing in a domain that includes complex-shaped objects. Unlike FFVHC-ACE, this study solves the governing equation of the space-charge density, formulated as partial differential equations. Additionally, a three-dimensional Poisson equation solver is formulated using the finite element method, and the global matrix for the entire computational domain is inverted using a direct method (LDLT decomposition). 

Another challenge is that plasma fluid models considering electrons with two or more components, which can predict the inception voltage of corona discharge in a closed form, inherently include multiple disparate time scales. For example, ions, which are much heavier than electrons, advect between electrodes on a time scale of milliseconds due to electric field drift, whereas the time scale of electron-impact ionization reactions is below a microsecond. Additionally, the Maxwell relaxation time scale can drop below a nanosecond, especially when the charge density is high.

Based on these findings, Shibata et al. \cite{shibata2016performance} applied perturbation methods to these equations, allowing an efficient and practically accurate determination of both the corona inception voltage and the thrust-to-power ratio. However, their approach required solving large-scale eigenvalue problems, limiting its applicability to two-dimensional cases. In this study, we extend their method with special modifications to expand its applicability to three-dimensional problems, which are crucial for practical applications. Their method reformulated the problem as an eigenvalue problem, solving it to determine the applied voltage and the corresponding eigenvector when the rightmost eigenvalue is positioned at the origin.

In general, finding the rightmost eigenvalue (the eigenvalue with the largest real part) of a very high-dimensional (e.g., 200 million-dimensional) real asymmetric matrix is extremely difficult in terms of computational cost. However, computing the rightmost eigenvalue of a real asymmetric matrix with at most 10,000 dimensions can be done in a short time even on a mid-range laptop. 

Therefore, in this study, we consider decomposing the problem into many smaller eigenvalue problems, enabling its application to practical, three-dimensional complex-shaped real-world problems, such as estimating the inception voltage of a solid-state electro-aerodynamic propulsion system.

Based on the above, the central academic question of this research is: ``Can practical solid-state electro-aerodynamic propulsion systems be efficiently analyzed with sufficient precision for practical applications?''. Therefore, the objective of this research is to establish a performance predictor capable of numerically analyzing fully three-dimensional solid-state electro-aerodynamic propulsion systems.
\begin{figure}
    \centering
    \includegraphics[width=0.75\linewidth]{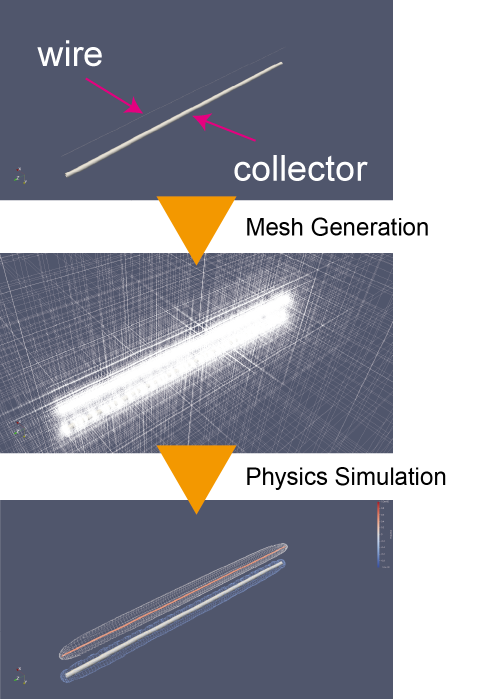}
    \caption{The procedures using the fully automatic grid generation. As shown in the first figure, a typical solid-state electro-aerodynamic propulsion system consists of a pair of electrodes (the wire and collector). As shown in the second figure, an unstructured mesh with a scale of one million nodes can be generated within a few seconds even on a laptop using the proposed method (although additional computation for inter-element connectivity is also required). The third figure shows the spatial distribution of the scalar electric potential, visualized using isosurfaces of equal potential.}
    \label{fig:procedures}
\end{figure}
\begin{figure}
    \centering
    \includegraphics[width = 0.7\linewidth]{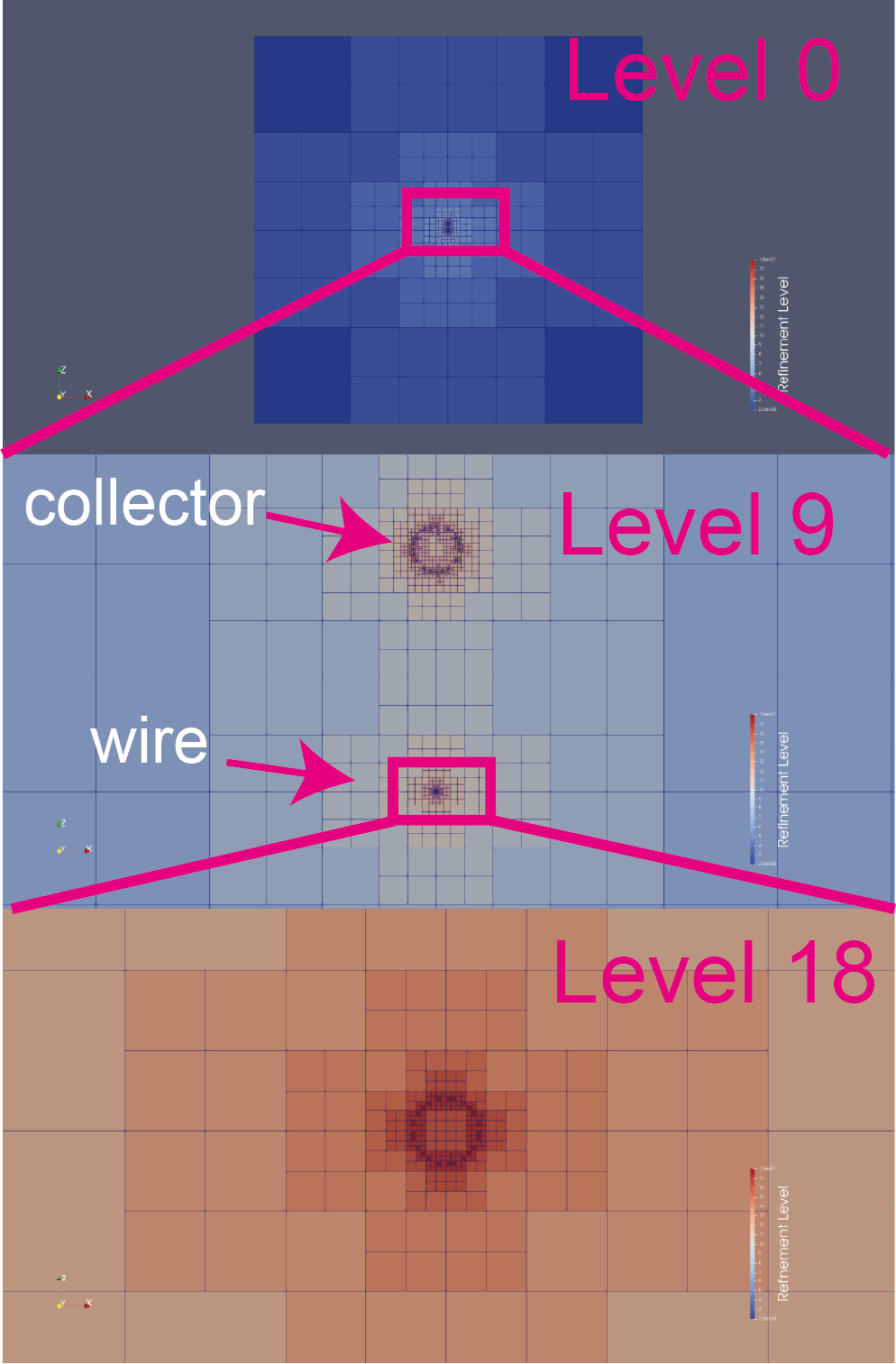}
    \caption{The multiscale nature of the solid-state electro-aerodynamic propulsion systems in space.}
    \label{fig:spatial_scales}
\end{figure}

Here, our code developed, ANEMONE is a C++ object-oriented code accelerated by OpenMP and Intel Math Kernel Library (MKL), combined with Eigen library. This code can analyze solid-state electro-aerodynamic propulsion systems of arbitrary electrode arrangement and shapes. The computational grid generation is fully automatically executed, on the basis of the hierarchical Cartesian grid method. 
ANEMONE are acronyms for [A]dvanced [N]umerical [E]lectroaerodynamic [M]odeling and [O]ptimization for [N]on-mechanical [E]ngines. The procedures with this new simulator are shown in Figure~\ref{fig:procedures}. Moreover, Figure~\ref{fig:spatial_scales} shows the multiscale nature of the solid-state electro-aerodynamic propulsion systems in space together with a hierarchical Cartesian grid example.

\section{Related works}
The representative performance metrics of a solid-state electro-aerodynamic propulsion system are the corona discharge onset voltage, the energy conversion efficiency, and the thrust density. However, there have been no studies that numerically analyze the performance of solid-state electro-aerodynamic propulsion systems in a fully three-dimensional space, and previous research has been limited to approximate analyzes in cases of two-dimensional \cite{xu2019higher, shibata2016performance, martins2011modeling, martins2011influence, martins2012simulation, martins2013modelling, picella2024numerical} or two-dimensional axisymmetric \cite{granados2017single, gomez2023model} cases.
In contrast, this study adopts a three-component plasma fluid model, enabling the analysis of fully three-dimensional solid-state electro-aerodynamic propulsion systems.

In 2015, Shibata et al. \cite{shibata2015global} applied the global linear stability analysis method to a three-component plasma fluid model (electrons, positive and negative ions) and its exact solution, the uncharged equilibrium solution. They successfully determined the voltage at which the uncharged equilibrium solution becomes unstable, which corresponds to the bifurcation and, therefore, the inception (or breakdown) voltage in the equations. In this method, it is typically necessary to solve a very large eigenvalue problem to determine the sign of the real part of the rightmost eigenvalue, which incurs a high computational cost. Related works are, e.g. \cite{almeida2020simple, almeida2024theory, benilov2021practical, suzuki2020bifurcation}.

Furthermore, Shibata et al. \cite{shibata2016performance} extended their method and successfully determined the thrust-to-current ratio $C_\eta$, a representative performance metric of solid-state electro-aerodynamic propulsion systems, using the perturbation method. However, the global matrix of the finite element method has a number of rows twice the number of nodes in the computational domain, and although theoretically possible, the application of this method to three-dimensional solid-state electro-aerodynamic propulsion systems was not straightforward.

Another contribution of the previous study was that they theoretically derived the following approximate law:
\begin{eqnarray}
    C_\eta := \frac{T}{I} \approx \frac{\delta T}{\delta I},
\end{eqnarray}
where $T$ is the total thrust, and $I$ is the total current of any solid-state electro-aerodynamic propulsion system. The quantities with the delta mean the perturbation on the uncharged solution. This law was originally derived by globally integrating the local components of the Maxwell stress tensor (MST) and the current density on the electrode surface \cite{shibata2016performance}. However, since no additional assumptions were imposed on the nature of the perturbation, it is naturally applicable to the perturbation corresponding to the earliest discharge path that arises along a single curve. In other words, the relation remains valid even when the perturbation lacks spatial extent and takes the form of an impulsive quantity in space.
However, for discharge paths corresponding to discharge voltages higher than the breakdown initiation voltage, there is no guarantee that the relation holds with high accuracy.
Guided by the above considerations, we continue to develop the methodology in the following sections.

\section{Method}
\subsection{The governing equations}
The three-component (i.e., electrons, positive and negative ions) plasma fluid model is given by the following drift-reaction equations:
\begin{eqnarray}
    \label{eqn:conservation_law_electrons}
    \frac{\partial \rho_e}{\partial t}  - \nabla \cdot \bm{J}_e &=& (\alpha - \eta) \mu_e E \rho_e, \\
     \label{eqn:conservation_law_p}
    \frac{\partial \rho_+}{\partial t}  + \nabla \cdot \bm{J}_+ &=& \alpha \mu_e E \rho_e,
\end{eqnarray}
and
\begin{equation}
    \frac{\partial \rho_-}{\partial t}  - \nabla \cdot \bm{J}_- = \eta \mu_e E \rho_e,
    \label{eqn:conservation_law_n}
\end{equation}
where the current density fluxes are given by 
\begin{equation}
\bm{J}_e = \mu_e \rho_e \bm{E},
\end{equation}
\begin{equation}
\bm{J}_+ = \mu_+ \rho_+ \bm{E},
\end{equation}
and
\begin{equation}
\bm{J}_- = \mu_- \rho_- \bm{E},
\end{equation}
where $\bm{E}$ is the electric field vector, and $E$ is the strength of the electric field vector.
$t$ is the time, $\alpha$ and $\eta$ are the electron impact ionization coefficient and the attachment coefficient as functions of the electric field strength (local field approximation).
$\mu_e$, $\mu_+ (=2.43\times 10^{-4}\ \mathrm{V\cdot s \cdot m^{-2}})$, $\mu_-$ are the mobility of electrons, positive and negative ions, respectively. $\rho_e$, $\rho_+$, and $\rho_-$ are the electron density and positive and negative ion densities. We assume atmospheric pressure air. All swarm parameters in this study were taken from \cite{morrow1997streamer}.

The corresponding boundary conditions are,
when the electric field is directed away from the electrode ($\bm{E} \cdot \hat{\bm{n}} > 0$, where $\hat{\bm{n}}$ is the outward normal vector of the electrode),
\begin{eqnarray}
    \bm{J}_+ \cdot \hat{\bm{n}} = 0,
\end{eqnarray}
and if not,
\begin{eqnarray}
    \gamma \cdot \bm{J}_+ \cdot \hat{\bm{n}} =  \bm{J}_e \cdot \hat{\bm{n}} ,
\end{eqnarray}
and
\begin{eqnarray}
    \bm{J}_- \cdot \hat{\bm{n}} = 0,    
\end{eqnarray}
where $\gamma (=1.0\times 10^{-3})$ is the secondary electron emission coefficient, approximated with a constant in this study. 

The model is solved with the Poisson equation for the scalar electric potential ($\phi$) to close the equation system,
\begin{equation}
    \nabla^2 \phi = -\frac{1}{\epsilon_0} \cdot \left(\rho_+ - \rho_- - \rho_e\right),
    \label{eqn:poisson}
\end{equation}
and 
\begin{equation}
    \bm{E} = -\nabla \phi,
    \label{eqn:efield}
\end{equation}
where $\epsilon_0 (=8.85 \times 10^{-12}\ \mathrm{F \cdot m^{-1}})$ is the permittivity of the air. We set the boundary condition for the Poisson equation as follows, on the anode and cathode electrodes,
\begin{eqnarray}
    \phi_{anode} &=& \frac{V_{apply}}{2}, \\
    \phi_{cathode} &=& -\frac{V_{apply}}{2},
\end{eqnarray}
and on the open boundaries,
\begin{eqnarray}
    \left( \nabla \phi \right) \cdot \hat{\bm{n}} = 0. 
\end{eqnarray}

\subsection{Transformation to the ordinal differential equations}
We define the new coordinate $s$ such that 
\begin{equation}
    \frac{d\bm{x}}{ds} = \bm{E},
\end{equation}
and using the chain-rule, we have
\begin{equation}
    \bm{E} \cdot \nabla \left(\mu_e \rho_e\right) = \frac{\partial (\partial\mu_e \rho_e)}{\partial s},
\end{equation}
\begin{equation}
    \bm{E} \cdot \nabla \left(\mu_+ \rho_+\right) = \frac{\partial (\partial\mu_+ \rho_+)}{\partial s},
\end{equation}
and
\begin{equation}
    \bm{E} \cdot \nabla \left(\mu_- \rho_- \right) = \frac{\partial (\partial\mu_- \rho_-)}{\partial s}.
\end{equation}

Therefore, we have
\begin{eqnarray}
    \label{eqn:conservation_law_electrons}
    \frac{\partial \rho_e}{\partial t}  - \frac{\partial (\mu_e \rho_e)}{\partial s} - \mu_e \rho_e \nabla \cdot \bm{E} &=& (\alpha - \eta) \mu_e E \rho_e, \\
    \label{eqn:conservation_law_p}
    \frac{\partial \rho_+}{\partial t}  + \frac{\partial (\mu_+\rho_+)}{\partial s} +\mu_+ \rho_+ \nabla \cdot \bm{E} &=& \alpha \mu_e E \rho_e,
\end{eqnarray}
and
\begin{equation}
    \frac{\partial \rho_-}{\partial t}  - \frac{\partial (\mu_-\rho_-)}{\partial s} - \mu_- \rho_- \nabla \cdot \bm{E} = \eta \mu_e E \rho_e.
    \label{eqn:conservation_law_n}
\end{equation}

\subsection{Transformation to the eigenvalue problem, and its integral form}
We linearize the governing equation around the uncharged equilibrium following \cite{shibata2015global},
\begin{equation}
    \rho_e = \bar{\rho}_e (=0) + \tilde{\rho}_e,
\end{equation}
\begin{equation}
    \rho_+ = \bar{\rho}_+ (=0) + \tilde{\rho}_+,
\end{equation}
and
\begin{equation}
    \rho_- = \bar{\rho}_- (=0) + \tilde{\rho}_-.
\end{equation}

The quantities with the bar symbol are evaluated on the Laplace field, that is, $\nabla^2 \bar{\phi} = 0$ and $\bar{\bm{E}} = - \nabla \bar{\phi}$.

We insert the above equations into the governing equations.
Then, we ignore all the second- or higher-order quantities. Therefore, we have
\begin{eqnarray}
    \label{eqn:conservation_law_electrons}
    \frac{\partial \tilde{\rho}_e}{\partial t}  - \frac{\partial (\bar{\mu_e} \tilde{\rho}_e)}{\partial s} &=& (\bar{\alpha} - \bar{\eta}) \bar{\mu}_e \bar{E} \tilde{\rho}_e, \\
    \label{eqn:conservation_law_p}
    \frac{\partial \tilde{\rho}_+}{\partial t}  + \frac{\partial (\bar{\mu}_+ \tilde{\rho}_+)}{\partial s} &=& \bar{\alpha} \bar{\mu}_e \bar{E} \tilde{\rho}_e,
\end{eqnarray}
and
\begin{equation}
    \frac{\partial \tilde{\rho}_-}{\partial t}  - \frac{\partial (\bar{\mu}_-\tilde{\rho}_-)}{\partial s} = \bar{\eta} \bar{\mu}_e \bar{E} \tilde{\rho}_e.
    \label{eqn:conservation_law_n}
\end{equation}

Furthermore, we put
\begin{equation}
    \tilde{\rho}_e (s, t) = \hat{\rho}_e (s) \exp{(\lambda t)},
\end{equation}
\begin{equation}
    \tilde{\rho}_+ (s, t) = \hat{\rho}_+ (s) \exp{(\lambda t)},
\end{equation}
\begin{equation}
    \tilde{\rho}_- (s, t) = \hat{\rho}_- (s) \exp{(\lambda t)},
\end{equation}
to obtain
\begin{eqnarray}
    \label{eqn:evp_e}
    \lambda \hat{\rho}_e &=& \frac{d(\bar{\mu_e} \hat{\rho}_e)}{ds}  + (\bar{\alpha} - \bar{\eta}) \bar{\mu}_e \bar{E} \hat{\rho}_e, \\
    \label{eqn:evp_p}
    \lambda \hat{\rho}_+ &=& -\frac{d(\bar{\mu}_+ \hat{\rho}_+)}{ds} + \bar{\alpha} \bar{\mu}_e \bar{E} \hat{\rho}_e,
\end{eqnarray}
and
\begin{equation}
    \lambda\hat{\rho}_- = \frac{d(\bar{\mu}_-\hat{\rho}_-)}{ds} + \bar{\eta} \bar{\mu}_e \bar{E} \hat{\rho}_e.
    \label{eqn:evp_n}
\end{equation}

The above three equations constitute an eigenvalue problem of the form together with the linearized boundary conditions:
\begin{equation}
    \lambda \bm{v} = J (V_{apply}) \bm{v},
\end{equation}
where $J$ is the Jacobian matrix with a parameter (the applied voltage), and $\bm{v}$ is the eigenfunction corresponded to the eigenvalue $\lambda$. The eigenvalue problem is obtained for each characteristic curve.

These many relatively small-scale eigenvalue problems obtained through decomposition can be numerically solved using an eigenvalue solver by first discretizing the problem in space. However, the parameter (the applied voltage) when the rightmost eigenvalue is located at the origin can actually be determined as the solution of a nonlinear integral equation. The eigenfunction corresponding to the rightmost eigenvalue can also be expressed in an integral form. We derive those equations.

Specifically, from Eq.~\ref{eqn:evp_e}, we can derive
\begin{equation}
    \bar{\mu_e} \hat{\rho}_e = C_e \exp{\left[- \int ( \bar{\alpha} - \bar{\eta}) \bar{E} ds\right]},
\end{equation}
where $C_e \not= 0$ is an integral constant corresponded to the degree of freedom for the eigenfunctions. We set $C_e = 1$ throughout this study.

Moreover, using the above equation, we can derive
\begin{equation}
\frac{d\bar{\mu}_+ \hat{\rho}_+}{ds} =  C_e \bar{\alpha} \bar{E} \exp{\left[- \int ( \bar{\alpha} - \bar{\eta}) \bar{E} ds\right]}.
\end{equation}

Therefore,
\begin{equation}
\bar{\mu}_+ \hat{\rho}_+ = C_e \int \left\{\bar{\alpha} \bar{E}\exp{\left[- \int ( \bar{\alpha} - \bar{\eta}) \bar{E} ds\right]} \right\} ds' + C_+
\end{equation}
Similarly,
\begin{equation}
\bar{\mu}_- \hat{\rho}_- = -C_e \int \left\{\bar{\eta} \bar{E} \exp{\left[- \int ( \bar{\alpha} - \bar{\eta}) \bar{E} ds\right]} \right\} ds' + C_-
\end{equation}
Those equations are the eigenfunctions corresponded to $\lambda = 0$.

Furthermore, we set $s=0$ as the anode and $s=L$ as the cathode.
Because $\hat{\rho}_+ = 0$ on the anode, we have $C_+ = 0$.
Because $\hat{\rho}_- = 0$ on the cathode, we have
\begin{equation}
    C_- = C_e \int_0^L \left\{\bar{\eta} \bar{E} \exp{\left[- \int_0^\sigma ( \bar{\alpha} - \bar{\eta}) \bar{E} d\sigma \right]} \right\} ds.
\end{equation}

The secondary electron emission boundary condition is implemented as 
\begin{equation}
    \gamma \bar{\mu}_+ \hat{\rho}_+ (s=L) = \bar{\mu}_e \hat{\rho}_e (s=L).
\end{equation}

Therefore,
\begin{eqnarray}
    \gamma \cdot \int_0^L \left\{\bar{\alpha}(E(s)) \bar{E}(s)\exp{\left[- \int_0^s ( \bar{\alpha}(E(s')) - \bar{\eta}(E(s'))) \bar{E}(s') ds' \right]} \right\} ds \nonumber \\
    = \exp{\left[- \int_0^L ( \bar{\alpha}(E(s')) - \bar{\eta}(E(s'))) \bar{E}(s') ds' \right]}
\end{eqnarray}

Here, we define
\begin{eqnarray}
    F(s) := \exp{\left[- \int_0^s ( \bar{\alpha}(E(s')) - \bar{\eta}(E(s'))) \bar{E}(s') ds' \right]}.
\end{eqnarray}

Using this definition, we have
\begin{eqnarray}
    \gamma\int_0^L \bar{\alpha}(s) \bar{E}(s) F(s) ds = F(L).
\end{eqnarray}

The root ($V_{apply}$) of this non-linear integral equation defines the discharge inception voltage ($V_{on}$).

\subsection{Numerical algorithm for solving the Laplace equation}
\label{subsec:poisson}
To solve the Laplace equation, which is an elliptic partial differential equation, and determine the spatial distribution of the scalar electric potential and the electric field vector, the three-dimensional Galerkin finite element method on 1-irregular meshes \cite{morton1995new} was adopted and implemented from scratch. In the hierarchical Cartesian grid method, hanging nodes inevitably appear between fine elements (cubes) and coarse elements, requiring special treatment. Note that the number of nodes per element is at least 8 but can be as many as 26.

The computational mesh generation is performed fully automatically based on the STL (Stereolithography) file representing the electrode shape, which is input into the simulator. The system is designed to input separate definition files for the anode and cathode. The process utilizes an octree nesting structure, allowing different refinement levels to be set for the surfaces of the anode and cathode electrodes.
The neighboring element search is performed efficiently using a KD tree without the need for a full element search. If the refinement level of a neighboring element is lower by two or more levels compared to the refinement level of the current element, the lower-level element is recursively refined.
As a result, the mesh is generated in such a way that the absolute difference in refinement levels between all adjacent elements is 0 or 1. This significantly simplifies the definition of shape functions in the finite-element method.
All elements are cubic in shape, but there is a large degree of freedom in their sizes and node arrangements. However, leveraging the fact that all elements are similar in shape, elements with the same node arrangement, regardless of scale, can store their corresponding local matrices in a cache. Using these cached data, the assembly of the global matrix has been significantly accelerated.

The grid used to represent the solid-state electro-aerodynamic propulsion system in this study is spatially multilevel, making the global matrix of the finite element method ``stiff''. Under such ill conditions, the linear solver PARDISO \cite{schenk2004solving} is adopted, as it enables robust matrix inversion.

Finally, the Laplace field is linear with respect to the applied voltage, so the scalar potential for any given applied voltage can be obtained very easily and quickly by scaling it by a constant factor.

\subsection{Computation of the power conversion efficiency}
The power conversion efficiency $\eta$ is given by
\begin{eqnarray}
    \eta = \frac{T \cdot U_{\infty}}{J \cdot V_{\mathrm{apply}}} \times 100 \%,
\end{eqnarray} 
where $U_\infty$ is the flight velocity, $J$ is the current density on one of the electric field lines.
Note that the values of $T$ and $J$ are generally different for each line.
Here,
\begin{eqnarray}
    \bm{T} = \int_0^L (\rho_+ - \rho_e  - \rho_-) \cdot \bm{E} dl,
\end{eqnarray}
and
\begin{eqnarray}
    J = \left(\mu_+ \rho_+ + |\mu_e| \rho_e + |\mu_-| \rho_-\right) \cdot \bm{E} \cdot \hat{\bm{n}}.
\end{eqnarray}
The current density $J$ must be conserved on the same on a line; hence, one can evaluate and represent $J$ at an arbitrary point on the line.

However, so far, we have not evaluated the real quantities $\bm{T}$ and $J$.
Instead, we know the perturbed quantities as follows:
\begin{eqnarray}
    \hat{\bm{T}} = \int_0^L (\hat{\rho}_+ - \hat{\rho}_e  - \hat{\rho}_-) \cdot \bar{\bm{E}} dl,
\end{eqnarray}
and
\begin{eqnarray}
    \hat{J} = \left(\bar{\mu}_+ \hat{\rho}_+ + |\bar{\mu}_e| \hat{\rho}_e + |\bar{\mu}_-| \hat{\rho}_-\right) \cdot \bar{\bm{E}} \cdot \hat{\bm{n}}.
\end{eqnarray}

Here, one of the authors derived the generalized version of the following law previously \cite{shibata2016performance}, and we adopt it in this study.
\begin{eqnarray}
    \frac{\hat{\bm{T}}}{\hat{J}} \approx \frac{\bm{T}}{J}.
\end{eqnarray}
In the following, we deal only with thrust as a scalar in the direction of the thrust axis, rather than as a vector.

\subsection{Overall procedure}
In summary, the overall procedure is as follows.
\begin{enumerate}
    \item Solve the Laplace equation by the finite element method (subsec. \ref{subsec:poisson}),
    \item Compute the electric field vector (three components),
    \item Compute the characteristic curves (electric field lines),
    \item Compute the discharge inception voltage (by e.g. bi-section method),
    \item Compute the eigenfunction,
    \item Compute the efficiency $\eta$.
\end{enumerate}

The number of characteristic curves to be tracked was set at 500, and their initial starting positions were randomly assigned to the surface of the anode electrodes. The integration of physical quantities along the characteristic curve was performed using the Dormand–Prince method, an explicit solver with adaptive step size control. The position along the curve (measured as the length of the curve from the starting point) was used as the independent parameter, and the electric field at each point was interpolated using shape functions of the finite element method. The integration continued until the characteristic curve reached the surface of one of the cathode electrode(s). We set $\eta = 0$ and neglected negative ions, because we consider positive corona discharge in this study.

In the ANEMONE simulator, the mesh convergence was tested automatically; we adopted the technique of the adaptive mesh refinement technique (h-refinement) \cite{berger1984adaptive} to generate meshes with lower errors. 

\subsection{Numerical analysis targets and setup}
\subsubsection{SST (MIT)}
\begin{figure}
    \centering
    \includegraphics[width=\linewidth]{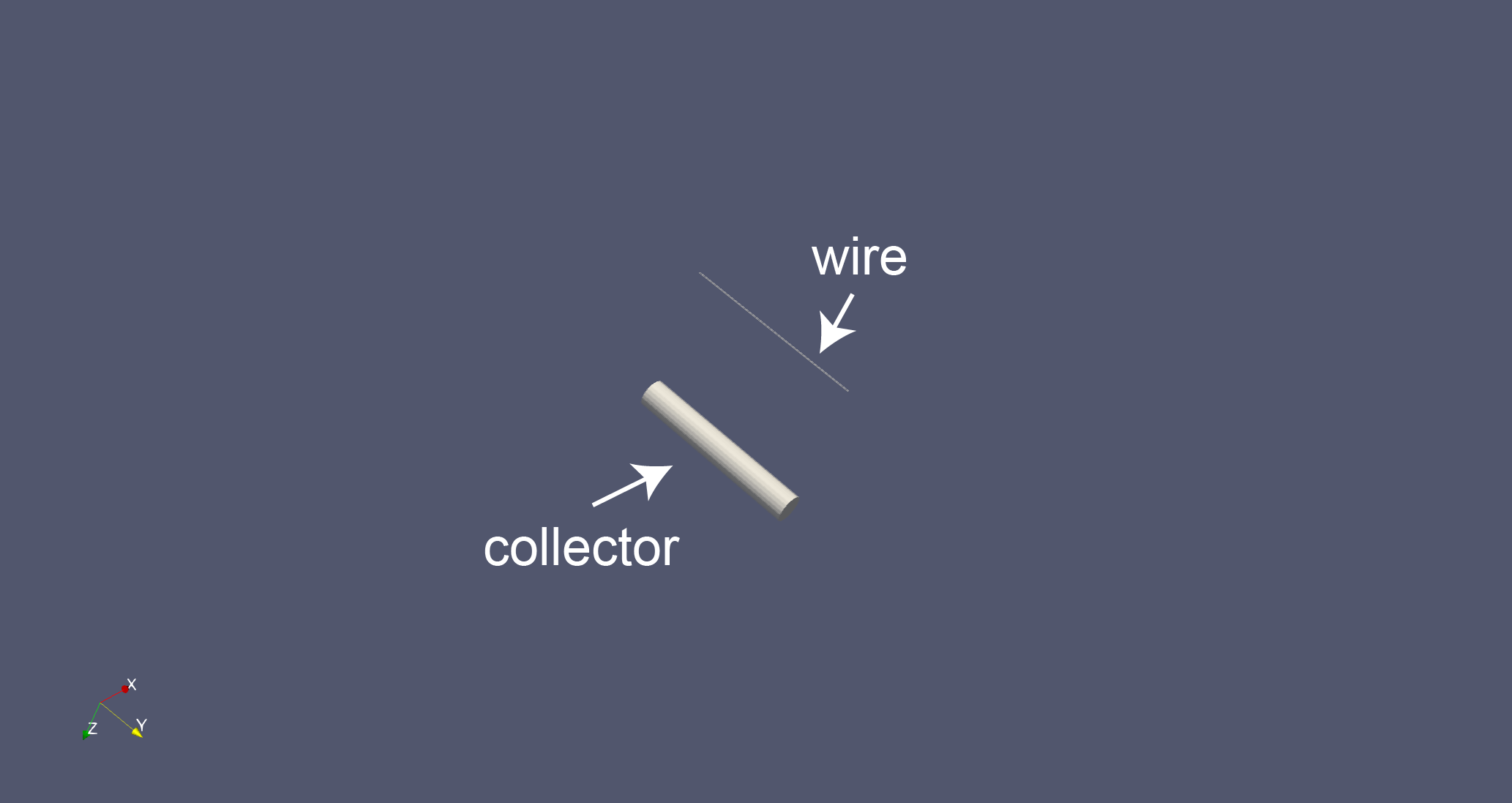}
    \caption{Overview of Single-Stage Thruster (SST).}
    \label{fig:sst-stl}
\end{figure}

First, to validate ANEMONE, we focus on solid-state electro-aerodynamic propulsion systems developed at MIT, namely SST (Single-Stage Thruster; Figure~\ref{fig:sst-stl}) \cite{masuyama2013performance}. The numerical results will be validated with the theoretical findings [e.g. \cite{masuyama2013performance, xu2019higher}]. Specifically, the corona inception voltage and the energy conversion efficiency are evaluated numerically and compared.

The SST electrodes consist only of a wire electrode with a small radius of curvature and a collector electrode with a relatively large radius of curvature. The diameter of the wire electrode is 202 $\mu$m, while the diameter of the collector electrode is 6.35 mm. The gap between the electrodes ($d$) was set at 3 cm. Since this is a three-dimensional analysis, the finite length in the depth direction must be set. To enable prediction with limited computational resources, particularly main memory, the actual thruster, which has a depth of 40 cm, is represented here with a reduced depth of 4 cm.

The computational domain is a cube with a side length of 2 m. This cube is set as subdivision level 0. The vicinity of the wire electrode is at level 15, and the vicinity of the collector electrode is at level 13. Mesh generation was completed within a few seconds on a laptop.

\subsubsection{HN-7 (Keio Univ.)}
\begin{figure}
    \centering
    \includegraphics[width=\linewidth]{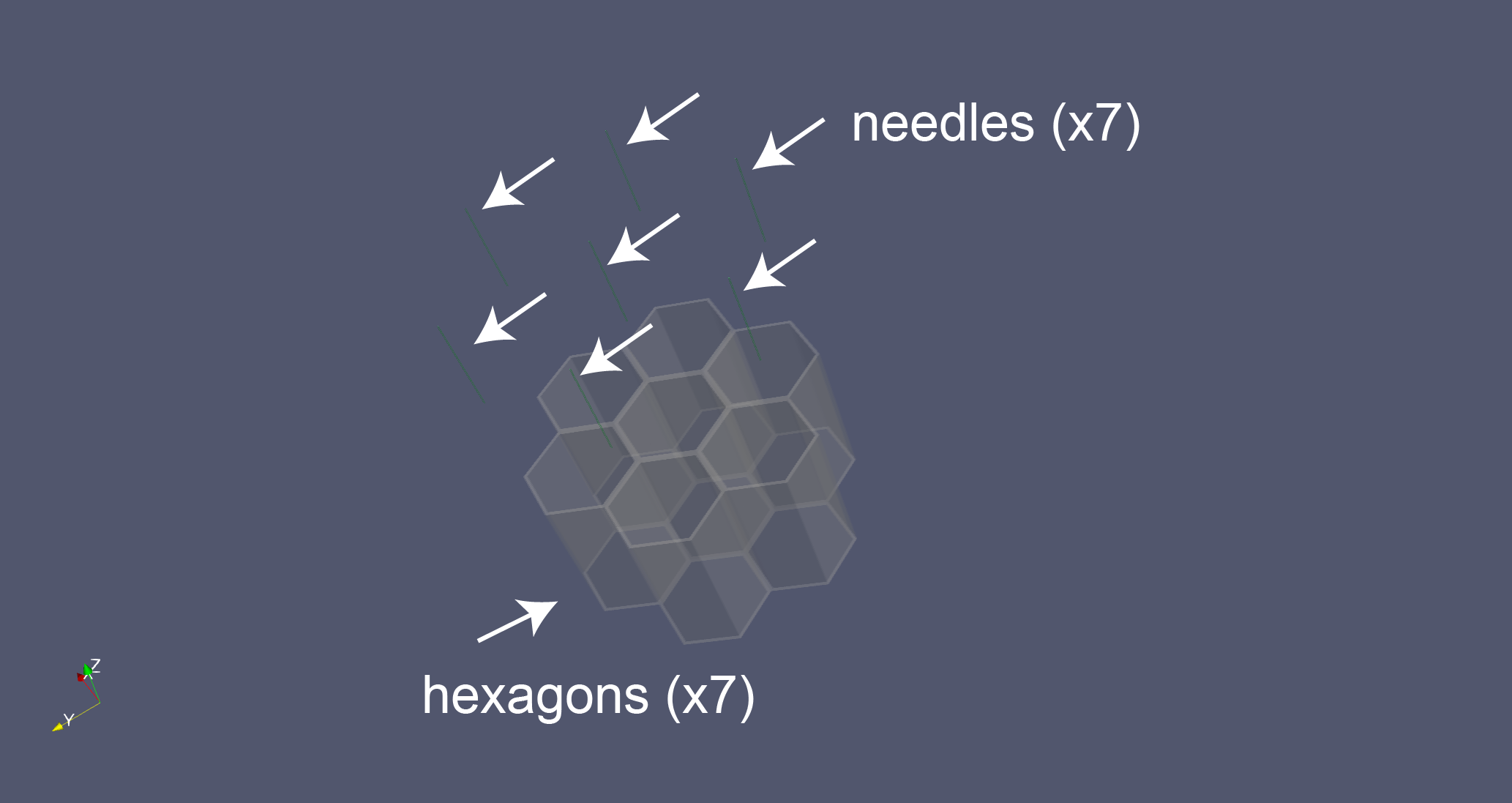}
    \caption{Overview of the HN-7 thruster.}
    \label{fig:hn-7}
\end{figure}

Second, we focus on HN-7 (a cluster of seven interconnected hexagons and a needle-shaped electrode; Figure~\ref{fig:hn-7}), developed at our institution \cite{katagiri2025}. Note that while the SST is a two-dimensional thruster, the HN-7 is essentially a three-dimensional thruster and cannot be analyzed without a three-dimensional solver.

As shown in Figure~\ref{fig:hn-7}, the HN-7 electrodes consist of seven needle electrodes with a small radius of curvature and hexagonal electrodes with a relatively large radius of curvature. The diameter of the needle electrodes is 200 $\mu$m, while the length and thickness of the hexagonal electrodes are 50 mm and 1 mm, respectively. The gap distance between the needle and the hexagonal electrodes ($d$) was set at 40 mm.

The computational domain is a cube with a side length of 1 m. This cube is set as subdivision level 0. The vicinity of the needle electrode is at level 15, and the vicinity of the hexagonal electrode is at level 12. Mesh generation was completed within a few seconds on a laptop.

\section{Results}
\begin{figure}
    \centering
    \includegraphics[width=\linewidth]{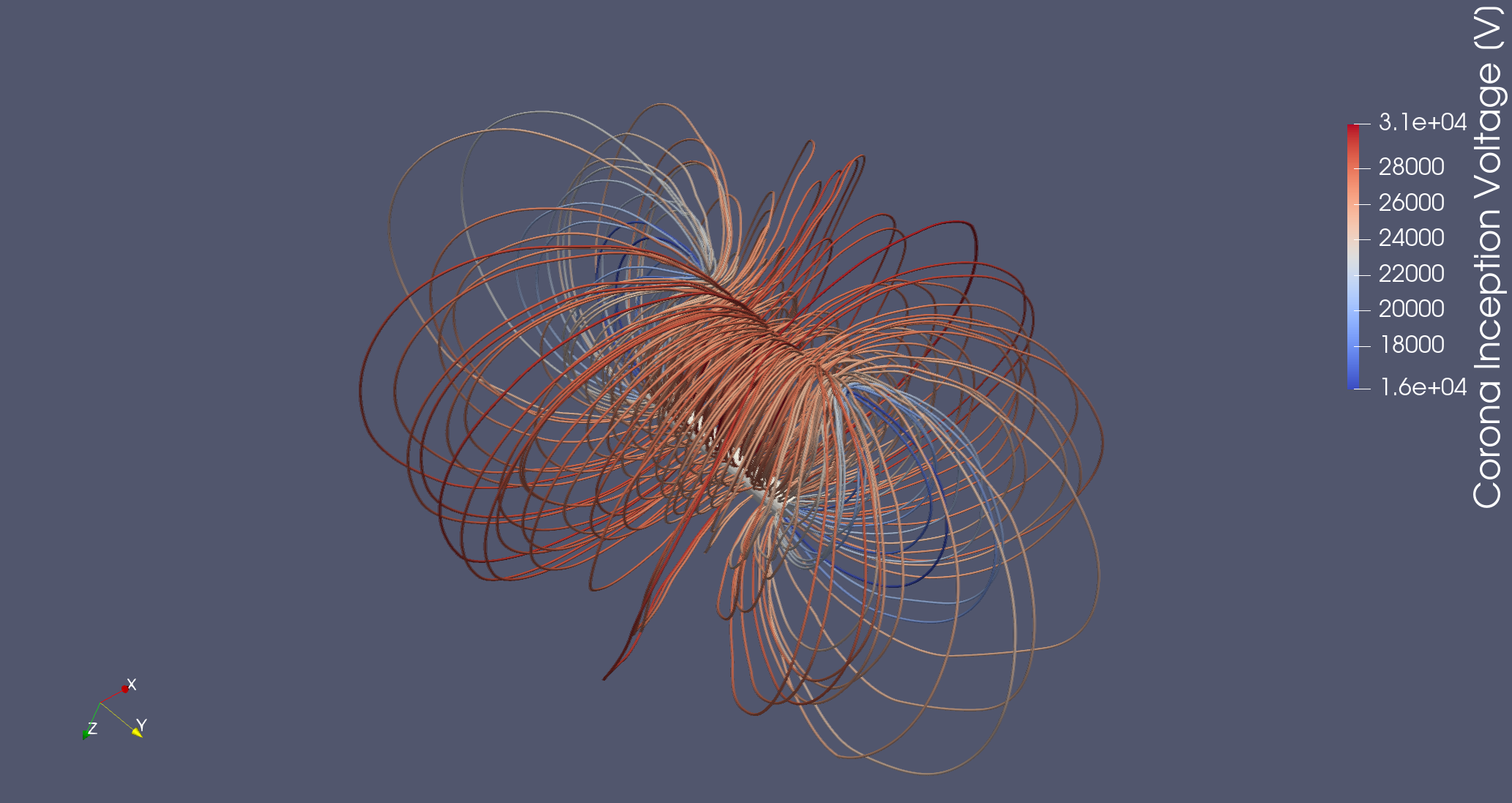}
    \caption{Corona inception voltage as a function of the electric field line (SST). Discharge begins at voltages below 16 kV in areas where the path is relatively short. As the discharge path extends, the corona discharge inception voltage corresponding to that path becomes relatively higher.}
    \label{fig:sst_civs}
\end{figure}
\begin{figure}
    \centering
    \includegraphics[width=\linewidth]{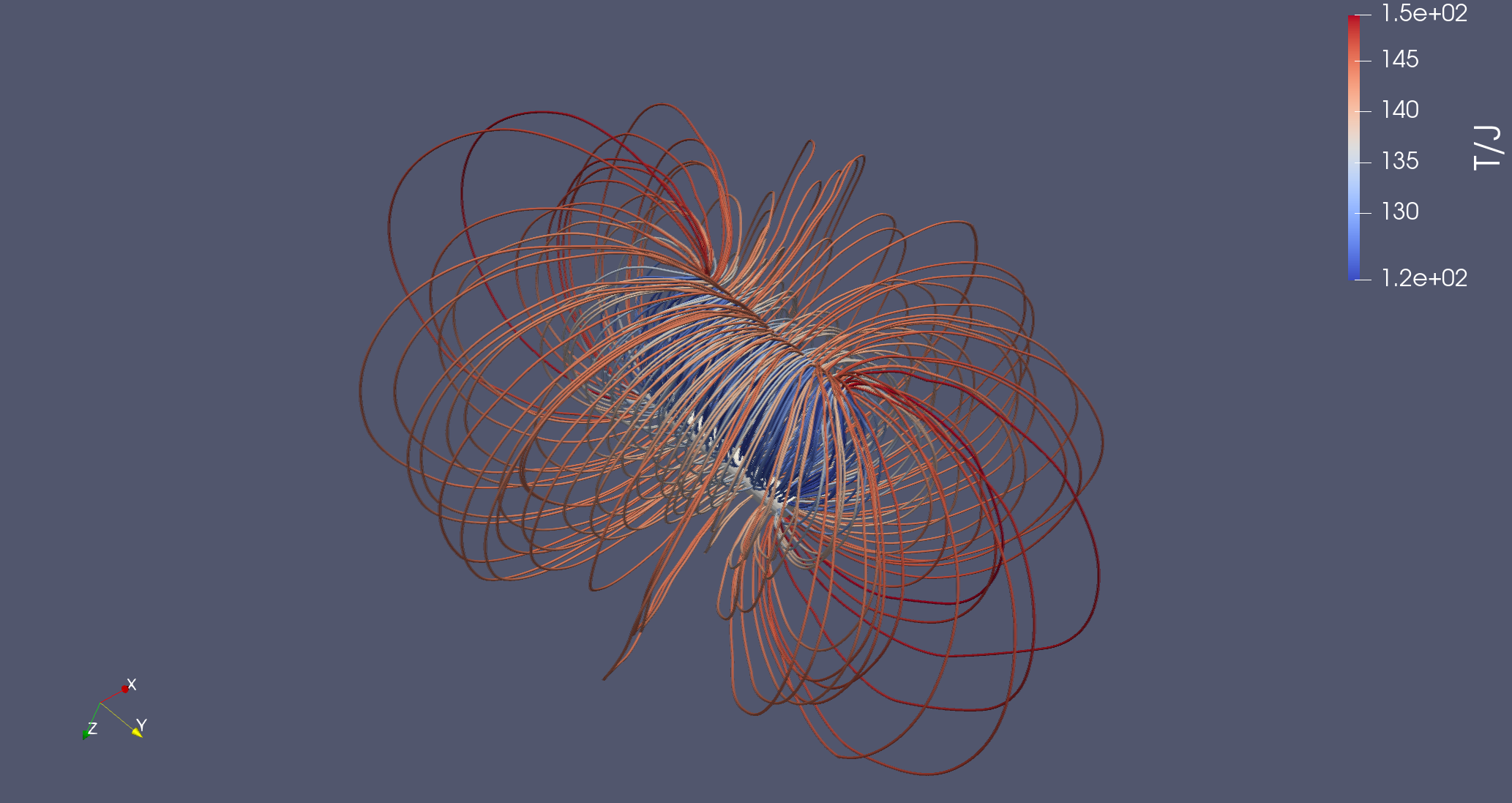}
    \caption{Thrust-to-current ratio as a function of the electric field line (SST).}
    \label{fig:sst_tbyjs}
\end{figure}

For the SST, Figure \ref{fig:sst_civs} shows the corona inception voltage as a function of the electric field line. Figure \ref{fig:sst_tbyjs} shows the thrust-to-current ratio as a function of the electric field line, with which we can calculate the energy conversion efficiency,

For the SST, the discharge begins at the path connecting the edge of the wire electrode and the edge of the collector electrode, and the corresponding corona inception voltage was estimated to be approximately 16 kV: an aspect characteristic of three-dimensional problems.

\begin{figure}
    \centering
    \includegraphics[width=\linewidth]{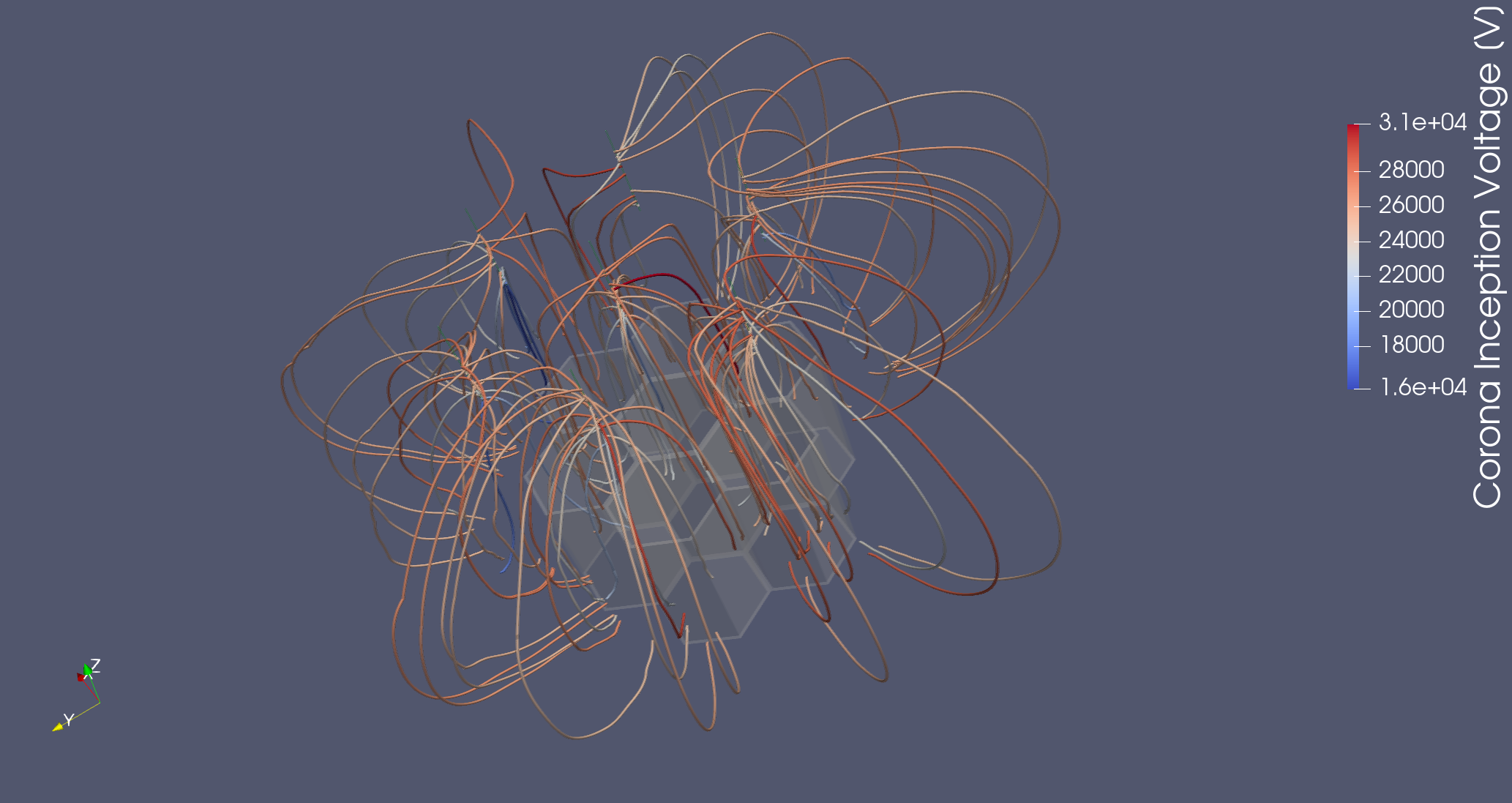}
    \caption{Corona inception voltage as a function of the electric field line (HN-7). Discharge begins at voltages below 16 kV in areas where the path is relatively short. As the discharge path extends, the corona discharge inception voltage corresponding to that path becomes relatively higher.}
    \label{fig:hn7_civs}
\end{figure}

\begin{figure}
    \centering
    \includegraphics[width=\linewidth]{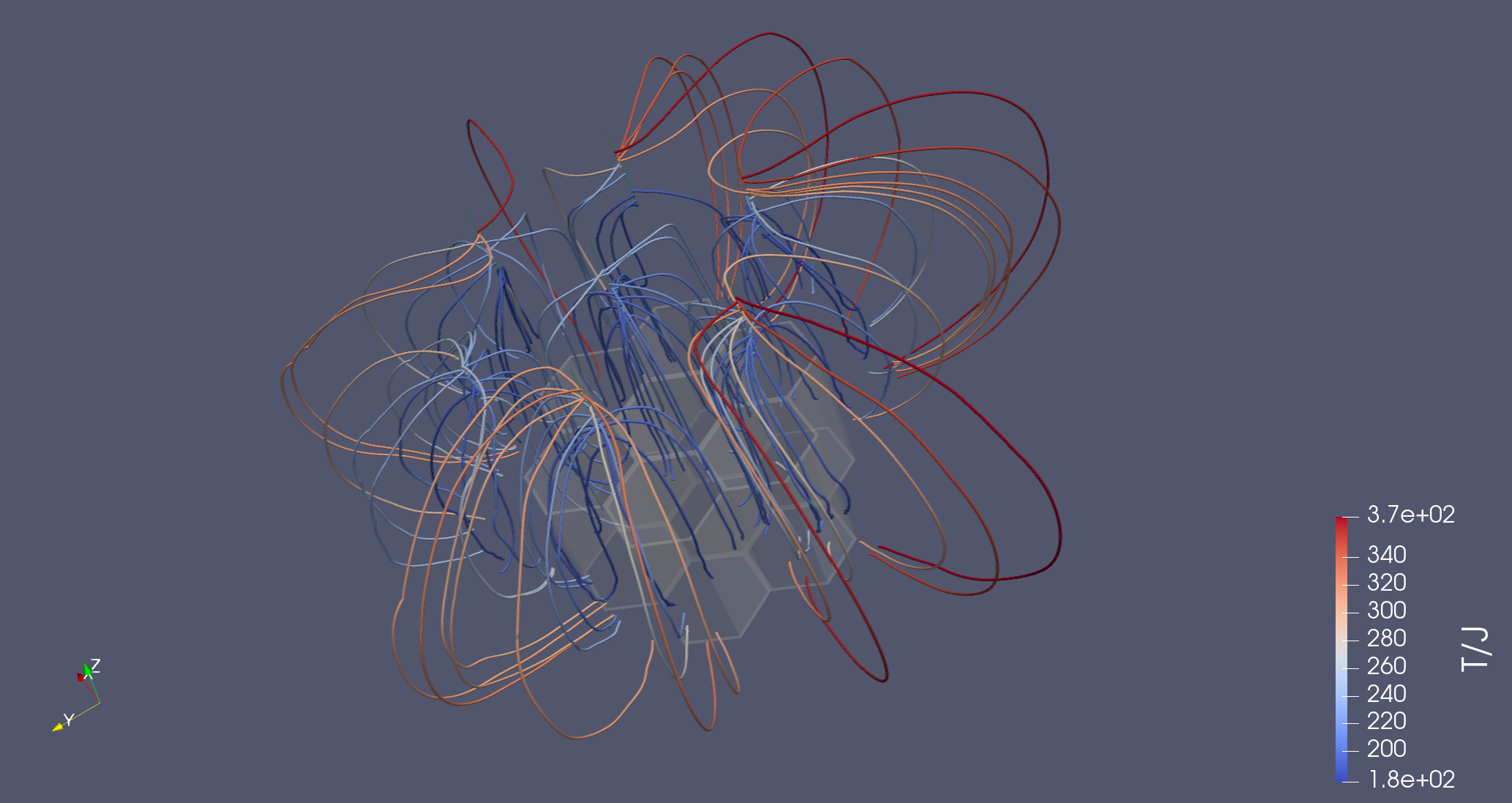}
    \caption{Thrust-to-current as a function of the electric field line (HN-7).}
    \label{fig:hn7_tbyjs}
\end{figure}

For HN-7, Figure \ref{fig:hn7_civs} shows the corona inception voltage as a function of the electric field line. Figure \ref{fig:hn7_tbyjs} shows the thrust-to-current ratio as a function of the electric field line.

\section{Discussion}
\subsection{Comparison with the theoretical and experimental results (SST)}
Theoretically \cite{christenson1967ion, masuyama2013performance}, the energy conversion efficiency of the SST with this setting is given by,
\begin{eqnarray}
    \eta &=& \frac{d}{\mu_+} \cdot \frac{U_\infty}{V_{apply}} \times 100 \\
    &=& \frac{0.03}{2.43 \times 10^{-4}} \cdot \frac{5.0}{4.0 \times 10^4} \times 100 \\
    &\sim& 1.5 \%,
\end{eqnarray}
where we assumed that the flight velocity $U_\infty$ of 5.0 m/s, and we used the applied voltage of 40 kV, which is much higher than the lowest voltage (16 kV) predicted with the ANEMONE simulator among a number of characteristic curves in this SST configuration. The efficiency along the path was calculated as 1.5 \%, indicating that ANEMONE can accurately predict performance.

\subsection{Performance evaluation of HN-7}
The path with the lowest discharge inception voltage (16 kV) originates from the lower tip of the needle electrode. In the following, we consider this path. Here, according to the theory \cite{christenson1967ion, masuyama2013performance}, the energy conversion efficiency of the HN-7 thruster is estimated by
\begin{eqnarray}
    \eta &=& \frac{d}{\mu_+} \cdot \frac{U_\infty}{V_{apply}} \times 100 \\
    &=& \frac{0.04}{2.43 \times 10^{-4}} \cdot \frac{5.0}{4.0 \times 10^4} \times 100 \\
    &\sim& 2.0 \%.
\end{eqnarray}
We assumed an operation voltage of 40 kV. However, the computed efficiency corresponding to the path of the lowest discharge onset voltage was 2.25 \%, which is higher than that of the theoretical value. As such, due to three-dimensional effects, a discrepancy from the theory could emerge.

\subsection{Novelty}
In fact, all previous studies dealing with solid-state electro-aerodynamic propulsion systems that utilize atmospheric pressure corona discharge have employed a single-component plasma fluid model. In particular, some previous studies \cite{shibata2016performance, shibata2022novel, granados2017single, granados2017study} have employed two- or multicomponent plasma fluid models, but their analyzes have been limited to the linear or weakly nonlinear regions, or have been conducted under atmospheric pressures approximately one-tenth of the standard atmospheric pressure.

As seen in Figure~\ref{fig:hn7_tbyjs}, the energy conversion efficiency is relatively higher in areas with longer discharge paths. This suggests that even though the corona discharge inception voltage increases with a longer path, the resulting increase in thrust or decrease in current density outweighs it.
Such inspection can contribute to the exploration of new discharge paths with higher energy conversion efficiency, for example, by locally suppressing the secondary electron emission coefficient along the initial discharge path (i.e., the path most prone to discharge).

\subsection{Advantages}
This makes it possible to visualize which part (electric field lines) of the discharge is reducing the overall energy conversion efficiency and to utilize this understanding of physical phenomena to improve performance.
Furthermore, with this method, the discharge inception voltage can be estimated without relying on empirical formulas, for example, Peek \cite{peek1929dielectric}. Therefore, it is considered that this method can be applied even to complex and specialized electrode geometries for which no empirical formulas exist.

\subsection{Limitations}
Discharges along various paths have been analyzed, but in reality, discharge begins at the location with the lowest inception voltage. As a result, space charge is generated, and to accurately analyze discharges along other paths, it is necessary to take into account the space charge from the initial discharge.

\subsection{Future works}
Here, instead of directly determining the thrust or thrust density itself, the efficiency of energy conversion was calculated from the electrical energy supplied to the thruster to the kinetic energy of the aircraft along each characteristic curve. However, for practical applications, it is also important to implement a code to calculate the thrust density and validate the results.

Using the feature of fully automated mesh generation, it is considered possible to perform optimization based on automatic exploration of electrode geometries to enhance the thrust density and energy conversion efficiency.

\section{Conclusion}
We succeeded in overcoming the challenge of spatial scales by employing a hierarchical Cartesian grid method, and the challenge of time scales by using the method of characteristics in conjunction with the perturbation method. Notably, the Laplace equation solver employs an adaptive mesh refinement method, while the characteristic curve solver utilizes adaptive step size control for spatial integration. In particular, the treatment of the Maxwell dielectric relaxation timescale was completely circumvented by considering the limit in which the charge density is zero throughout the entire space. The ANEMONE simulator for the design of solid-state electro-aerodynamic propulsion systems enables efficient predictive calculations of the corona inception voltage and the energy conversion efficiency immediately after discharge initiation along multiple electric field lines, making it possible to identify which characteristic curves contribute to the overall system performance well in practical three-dimensional settings.

\section*{Acknowledgement}
The simulator developed in this study (ANEMONE) was implemented from scratch by the authors without referring to the FFVHC-ACE code and has no affiliation with the Japan Aerospace Exploration Agency (JAXA).
This work was supported by the New Energy and Industrial Technology Development Organization (NEDO) project (Japan-U.S. International Joint Research and Development of a Solid-State electro-aerodynamic Propulsion System Using High Voltage).





\bibliographystyle{elsarticle-num}
\bibliography{main}







\end{document}